**A quantum computing approach to beam angle optimization**

Nimita Shinde[1*], Ya-Nan Zhu[1], Haozheng Shen[2], and Hao Gao[1*]

[1]Department of Radiation Oncology, University of Kansas Medical Center, USA

[2]University of Michigan-Shanghai Jiao Tong University Joint Institute, Shanghai Jiao Tong University, Shanghai, China

[*] Corresponding authors:

Nimita Shinde, and Hao Gao, Department of Radiation Oncology, University of Kansas Medical Center, USA.

Email addresses: nshinde@kumc.edu, hgao2@kumc.edu

**Abstract.**

**Background:** Beam angle optimization (BAO) is a critical component of radiation therapy (RT) treatment planning, where small changes in beam configuration can significantly impact treatment quality, especially for proton RT. Mathematically, BAO is a mixed integer programming (MIP) problem, which is NP-hard due to its exponential growing search space. Traditional optimization techniques often struggle with computational efficiency, necessitating the development of novel approaches.

**Purpose:** This study introduces QC-BAO, a hybrid quantum-classical approach that leverages quantum computing to solve the MIP formulation of BAO.

**Methods:** The proposed approach, QC-BAO, models BAO as an MIP problem, incorporating binary variables for beam angle selection and continuous variables for optimizing spot intensities for proton therapy. The proposed approach employs a hybrid quantum-classical framework, utilizing quantum computing to solve the binary decision component while integrating classical optimization techniques, including iterative convex relaxation and the alternating direction method of multipliers.

**Results:** Computational experiments were conducted on clinical test cases to evaluate QC-BAO's performance against clinically verified angles and a heuristic approaches, GS-BAO and AG-BAO. QC-BAO demonstrated improved treatment plan quality over clinical, GS-BAO and AG-BAO-selected angles. QC-BAO consistently increased the conformity index (CI) for target coverage while reducing mean and maximum doses to organs-at-risk (OAR). For instance, in the lung case, QC-BAO achieved a CI of 0.89, compared to 0.89 (clinical), 0.76 (GS-BAO) and 0.86 (AG-BAO), while lowering the mean lung dose to 2.78 Gy from 3.36 Gy (clinical), 4.80 Gy (GS-BAO) and 3.03 Gy (AG-BAO). Additionally, QC-BAO produced the lowest objective function values, confirming its superior optimization capability.

**Conclusions:** The findings highlight the potential of quantum-inspired algorithm to enhance the solution to BAO problem by demonstrated improvement in plan quality using the proposed method, QC-BAO. This study paves the way for future clinical implementation of quantum-accelerated optimization in RT.

## 1. Introduction

Radiation therapy (RT) plays a critical role in cancer treatment by delivering targeted radiation doses using multiple beam angles to tumors while minimizing exposure to healthy tissues and organs-at-risk (OAR). Beam angle optimization (BAO) [1] is a key aspect of RT treatment planning, selecting the optimal set of beam angles to achieve the best possible dose distribution. The selection of beam angles significantly impacts treatment plan quality in both intensity-modulated radiation therapy (IMRT) [2] and intensity-modulated proton therapy (IMPT) [3], as these plans often consist of only a few beam angles, and small changes in configuration can lead to notable differences in treatment outcomes [4].

Mathematically, BAO is a mixed-integer programming (MIP) [5,6] problem involving binary (0 or 1) variables for beam angle selection and continuous variables for optimizing spot intensities. However, BAO is a challenging combinatorial problem, with the number of possible beam angle combinations growing exponentially, leading to increasing computational complexity. Thus, BAO is classified as an NP-hard problem [7] and solving MIP formulation of BAO is computationally demanding due to its high-dimensional search space and complex constraints. Classical optimization methods, such as branch-and-bound algorithms [8], cutting-plane methods [9], and decomposition techniques [10], often struggle with efficiency.

To address these challenges, various methods have been proposed to tackle BAO, including heuristic and metaheuristic approaches [1,11-19], as well as group sparsity regularization techniques [20-23]. While these methods often offer computational efficiency, they often yield suboptimal solutions. Thus, there is a need for improved approaches to achieve near-optimal beam angle selection efficiently.

Quantum computing (QC) [24,25] presents a promising alternative to solve MIP. By leveraging quantum parallelism, QC can, in principle, leverage quantum parallelism to explore large combinatorial solution spaces efficiently. Recent advancements in variational quantum algorithms and gate-based quantum solvers [26] have demonstrated potential for solving binary optimization problems, making them well-aligned with the discrete nature of BAO. Recent advancements in QC have shown promise in radiation therapy optimization (see Section 4), and the increasing availability of quantum hardware, such as D-Wave's quantum annealers [27] and IBM's quantum processors [28], further supports its application in RT. To the best of our knowledge, quantum-inspired optimization techniques have not yet been applied to solve the BAO problem.

This work introduces a novel quantum-classical hybrid approach, QC-BAO, that formulates BAO as an MIP and solves its binary component using a quantum-inspired algorithm. Importantly, we implement the quantum algorithm using a simulator on classical hardware, not physical quantum processors. This allows us to evaluate quantum optimization strategies while remaining within currently accessible computational infrastructure. Our approach integrates classical optimization (for spot intensity optimization) with a quantum-inspired technique for binary beam angle selection, enabling efficient and scalable treatment planning. To our knowledge, this is the first application of quantum optimization techniques, even in simulated form, to the BAO problem. Computational experiments on clinical test cases demonstrate the effectiveness of this approach, indicating its potential for clinical implementation.

## 2. Methods

*2.1. MIP model for beam angle optimization*

For the problem of optimally selecting $N$ angles from a set of $B$ available beam angles, the proposed MIP optimization problem is

$$
\begin{aligned}
&\min_{x,y}\ f(d) \\
&s.t.\ d = \sum_{i \in B} y_i \left(A^i x^i\right) \\
&\quad x^i \in \{0\} \cup [G, +\infty\} \ \forall i \in B \\
&\quad \sum_{i \in B} y_i = N \\
&\quad y_i \in \{0,1\} \ \forall i \in B.
\end{aligned}
\tag{1}
$$

In Eq. (1), $A_i$ is the dose influence matrix for each beam angle $i \in B$, and $G$ denotes the minimum monitor unit (MMU) threshold value. The continuous decision variable $x^i$ corresponds to the spot intensity vector for each beam angle $i = 1, \ldots, B$ to be optimized, and the binary decision variable $y_i$ determines the beam angles selected in the plan. The first constraint in Eq. (1) defines the dose distribution $d$ based on the selected beam angles $y_i$ and their corresponding spot intensity vectors $x^i$. The second constraint enforces the MMU requirement [29,30], ensuring the treatment plan deliverability. The third and fourth constraints together impose a binary restriction on $y_i$ and enforce the selection of exactly $N$ beam angles from the available $B$ angles.

The first term in the objective function, $f(d)$, is defined as

$$
f(d) = \sum_{i=1}^{N_1} \frac{\mathrm{w}_1}{n_i} ||d_{\Omega_{1i}} - b_{1i}||_2^2 + \sum_{i=1}^{N_2} \frac{\mathrm{w}_2}{n_i} ||d_{\Omega_{2i}} - b_{2i}||_2^2 + \frac{\mathrm{w}_3}{n} ||\mathrm{d}_{\Omega_3} - b_3||_2^2. \tag{2}
$$

The three components of the objective function $f(d)$ are described below.

- **Target dose matching:** The first term represents $N_1$ least square error terms that measure the difference between the actual dose $d_{\Omega_{1i}}$ and the prescribed dose $b_{1i}$ for the target and OAR. Here, $\Omega_{1i}$ is the set of active indices, i.e., the set of voxel indices at which the actual dose differs from the prescribed dose value.
- **DVH-max constraint for OAR:** The second term in $f(d)$ incorporates $N_2$ dose volume histogram (DVH)-max constraints [31,32] for OAR. A DVH-max constraint ensures that at most a $p$ fraction of the total voxels in a given OAR receive a dose exceeding $b_{2i}$. To enforce this, $\Omega_{2i}$ is defined as the set of indices violating the constraint. Let $d'$ denote the dose $d$ sorted in descending order and let $n_i$ be the number of voxels in OAR $i$. Then, $\Omega_{2i} = \{j | j \geq p \times n_i\}$ if $d'_{p \times n_i} \geq b_{2i}$. Thus, if the DVH-max constraint is violated, the second term in $f(d)$ minimizes the least square error between the dose $d_{\Omega_{2i}}$, and the DVH-max dose $b_{2i}$.

- **DVH-min constraint for the target:** The third term enforces a DVH-min constraint [31,32] for the target, ensuring that at least a fraction $p$ of the total target voxels receive a dose greater than $b_3$. Again, sorting the dose $d$ in descending order as $d'$ and letting $n$ be the total number of voxels in the target, the active index set, $\Omega_3$, for DVH-min constraint is defined as $\Omega_3 = \{j | j \leq p \times n\}$ if $d'_{p\times n} \leq b_3$. If the DVH-min constraint is violated, the third term in $f(d)$ minimizes the least square error between the actual dose $d_{\Omega_3}$ at the violating indices and the minimum required dose $b_3$.

The solution to Eq. (1) is (i) a binary vector indicating the selected beam angles for dose delivery, and (ii) spot intensities corresponding to the selected beam angles. Section 2.2 presents an optimization method to solve the MIP formulation, Eq. (1), of the beam angle optimization problem.

### *2.2 QC-BAO method for beam angle optimization*

The proposed QC-BAO method to solve the beam angle optimization problem follows a two-step process:

1. **Step 1: Beam angle selection and spot intensity optimization:** In the first step, Eq. (1) is solved to determine the optimal beam angle selection. A quantum optimization algorithm implemented using a simulator on a classical machine is employed to optimize the binary variables $y_i$, which represent the selection of beam angles. The details of the algorithm used to solve Eq. (1) are provided in Sections 2.3 and 2.4. This step also yields the corresponding spot intensities for the selected beam angles. However, since the primary focus here is on beam angle selection, there remains an opportunity to further refine the spot intensities.
2. **Step 2: Additional spot intensity refinement:** In the second step, Eq. (1) is re-solved with the beam angles fixed according to the $y_i$'s determined in Step 1. This reduces the problem into a continuous optimization problem focused solely on refining the spot intensities. By isolating the continuous variables, this step enables a more precise adjustment of dose distribution for the selected beam angles.

### *2.3 Solution methodology for QC-BAO method*

*Step 1 of QC-BAO:* The first step of the QC-BAO method involves solving Eq. (1). However, Eq. (1) contains non-convex constraints, making it difficult to solve directly using standard optimization techniques. To enable more tractable optimization, auxiliary variables  are introduced, transforming the problem into a more structured form. This leads to the following reformulated problem

$$
\begin{aligned}
&\min_{x,y,z} \ f(d) \\
&s.t.\ \ d = \sum_{i \in B} y_i\left(A^i x^i\right) \\
&\qquad x^i = z^i \ \ \forall i \in B \\
&\qquad z^i \ \in \{0\} \ \cup [G, +\infty\} \ \ \forall i \in B \\
&\qquad \sum_{i \in B} y_i = \ N \\
&\qquad y_i \in \{0,1\} \ \ \forall i \in B.
\end{aligned} \tag{3}
$$

Next, the augmented Lagrangian of Eq. (3) is defined as

$$
\begin{aligned}
&\min_{x} \ f\left(\sum_{i \in B} y_i\left(A^i x^i\right)\right) + \frac{\mu_1}{2} \sum_{i \in B} ||x^i - z^i + \lambda_{1i}||_2^2 + \frac{\mu_2}{2}\left(\sum_{i \in B} y_i - N + \lambda_2\right)^2 \\
&s.t. \quad z^i \ \in \{0\} \ \cup [G, +\infty\} \ \ \forall i \in B \\
&\qquad\ \ y_i \in \{0,1\} \ \ \forall i \in B.
\end{aligned} \tag{4}
$$

Eq. (4) is solved using iterative convex relaxation (ICR) [33,34] and alternating direction method of multipliers (ADMM) [35,36], both of which have been successfully applied to inverse optimization problems [37-45]. The iterative method updates the active index sets for the DVH constraints. This is followed by sequential updates of each decision variable while keeping the others fixed.

A quantum-inspired optimization algorithm [24,25] is used to solve the binary subproblem for updating the beam selection variables $y_i$ at each iteration of the proposed method. Although the algorithm is designed for quantum hardware, we simulate it on classical hardware using a MATLAB-based solver. This simulation does not exploit true quantum parallelism but retains advantages such as stochastic sampling and efficient exploration of non-convex binary spaces. This quantum-inspired step is integrated into an iterative framework that also employs classical methods for updating continuous variables (spot intensities), resulting in a hybrid classical-quantum scheme [26,46]. An outline of this iterative strategy for solving Eq. (4) is provided in Algorithm 1. The first phase of the algorithm performs approximately 10 alternating updates between the continuous and binary variables, using the quantum-inspired solver to optimize $y_i$. A detailed explanation of this procedure is included in Section 2.4.

*Step 2 of QC-BAO:* Once a near-optimal set of beam angles is determined in the first step, Eq. (4) is solved again with $y_i$ fixed. The objective of this step is to further optimize the spot intensities for the selected beam angles. The solution methodology is similar to that outlined in Algorithm 1, however, in this step, $y_i$ remains fixed.

### *2.4 Updating primal variables in Algorithm 1*

A detailed explanation of Step 4b of Algorithm 1 is provided in this subsection.

1. Updating $x^i$: For each $i = 1, \dots, B$, fix all variables except $x^i$ in Eq. (4). Since the resulting minimization problem is unconstrained in $x^i$, solving it involves taking the first-order derivative of the objective function with respect to $x^i$ and setting it to zero. The value of $x^i$ is then obtained by solving the resulting linear system of equations.
2. Updating $z^i$: For each $i = 1, \dots, B$, fix all variables except $z^i$ in Eq. (4). The resulting minimization problem has a closed form solution, which is determined using soft thresholding:
$$z^i = \begin{cases} max\,(G, x^i - \lambda_{1i}), & if\ x^i - \lambda_{1i} \geq G/2 \\ 0, & otherwise. \end{cases}$$
3. Updating $y_i$: In Eq. (4), fix all variables except $y_i$. This reduces the subproblem to a quadratic unconstrained binary optimization (QUBO) [47], a problem well-studied in quantum optimization literature. Although QUBO problems can also be solved using classical MIP algorithms, in this work, QUBO is solved using a MATLAB solver called 'qubo', which simulates the behavior of a quantum optimization routine. While this does not leverage the full benefits of quantum computing, it allows for enhanced exploration of discrete solution spaces due to its stochastic sampling mechanism. The result is an effective, quantum-inspired method for BAO that operates entirely on classical hardware.

**Algorithm 1: Optimization method for solving Eq. (4)**

1. **Input:** Choose parameters $\mu_1, \mu_2, w_1, w_2, w_3$
2. **Initialization:** Randomly initialize $x, y$. Choose number of iterations $T$.
3. Set $z^i = x^i$, $\lambda_{1i} = \lambda_2 = 0$.
4. For $t = 1, \dots, T$
   a. **Identify active index sets:** Determine the active index sets $\Omega_{1i}, \Omega_{2i}, \Omega_3$ for the DVH constraints, as outlined in Section 2.1.
   b. **Update primal variables:** Sequentially update the primal variables $x^i, z^i, y_i$ by fixing all other variables and solving the resulting minimization problem for each.
   c. **Update dual variables:** Perform the following updates:
$$\lambda_{1i} = \lambda_{1i} + z^i - x^i$$
$$\lambda_2 = \lambda_2 + \sum_{i \in B} y_i - N.$$
5. **Output:** $x, y$

*2.5 Materials*

The binary component (QUBO) of the QC-BAO method is solved using quantum simulator on classical hardware. Though certain advantages of quantum hardware such as quantum parallelism are lost due to such simulation, it retains the advantage of enhanced exploration of non-convex binary spaces through quantum-inspired sampling mechanisms. These capabilities improve the likelihood of escaping local optima compared to purely classical solvers. The effectiveness of the proposed QC-BAO method is demonstrated by conducting experiments on three clinically relevant test cases: head-and-neck (HN) (8 Gy x 5 fractions), abdomen (6 Gy x 4 fractions), and lung case (2 Gy x 10 fractions). For each case, the computational efficiency and the resulting plan quality is compared against four benchmark methods.

- **Clinically verified angles**: These serve as a practical baseline. Beam directions in the clinical plans were chosen based on clinical experience, without computational optimization. Specifically, the HN case uses four coplanar beams at gantry angles (45º, 135º, 225º, 315º), while the abdomen and lung cases use three coplanar beams at (0º, 120º, 240º), all with a 0º couch angle. These angles represent typical treatment configurations used in actual patient treatments and reflect the performance achievable without BAO algorithms.
- **GS-BAO (Group Sparsity BAO) [23]**: A heuristic method that incorporates group sparsity regularization to implicitly select beam angles by penalizing unused beams in the fluence optimization problem. GS-BAO is relatively fast and computationally efficient, making it suitable for time-sensitive applications. It serves as a benchmark for evaluating the trade-off between speed and plan quality.
- **AG-BAO (Angle Generation BAO) [23]**: A heuristic method known for producing high-quality solutions through an iterative angle refinement process. AG-BAO typically yields better plans than GS-BAO but requires significantly more computation time. This method provides a performance ceiling for heuristic approaches, against which the accuracy of QC-BAO and MIP-based methods can be judged.
- **Classical MIP method**: The binary subproblem in the MIP formulation of the BAO problem is solved using classical MIP methods. Because the formulation is nonconvex, a genetic algorithm is used to solve the binary component. This approach allows for direct comparison with QC-BAO to understand whether quantum-inspired methods offer advantages in speed or optimality over purely classical global optimization.

For GS-BAO, AG-BAO, classical MIP and QC-BAO methods, beam selection is performed from a set of 72 non-coplanar angles, comprising 24 equally spaced gantry angles for each of three couch angles (0º, 30º, and 60º) yielding $N = 72$ in Eq. (1). The number of selected beam angles is fixed at $B = 4$ for the HN case and $B = 3$ for both the abdomen and lung cases.

The dose influence matrix is generated using MatRad [48] with a spot width of 5 mm on a 3 mm³ dose grid. CTV-based planning is performed with clinically defined constraints for all test cases. All plans are normalized to ensure that 95% of the target region receives at least 100% of the prescribed dose.

Plan quality is assessed using the following metrics: (a) conformity index (CI), (b) maximum dose delivered to tumor ($D_{max}$), (c) mean and max doses delivered to OAR. CI is defined as $V_{100}^2/(V \times V'_{100})$, where $V_{100}$ is the target volume receiving at least 100% of the prescription dose, $V$ is the total target volume, and $V'_{100}$ is the total volume receiving at least 100% of the prescription dose. The normalized maximum dose $D_{max}$ is calculated as $(D/D_p)$x100%, where D is the maximum dose delivered to the tumor, and $D_p$ is the prescription dose.

## 3. Results

### *3.1 Optimality of QC-BAO*

As shown in Table 1, QC-BAO achieves the lowest objective function value across all three clinical test cases, consistently outperforming all baseline methods. In the lung case, for example, QC-BAO attains an objective value of 6.62, outperforming AG-BAO (7.22), classical MIP (7.15), and GS-BAO (12.53). Similarly, in the abdomen and HN cases, QC-BAO yields objective values of 0.13 and 3.29, respectively, again outperforming the next-best method in each case. These results demonstrate that QC-BAO is effective at identifying near-optimal beam configurations. Furthermore, AG-BAO has been previously shown to produce near-optimal solutions using angle generation techniques (see Figure 1 in [23]). Table 1 shows that QC-BAO improves on the AG-BAO's result, demonstrating the near-optimality of the proposed method.

### *3.2 Computational efficiency and trade-offs*

In addition to improved optimization results, QC-BAO exhibits favorable computational performance. AG-BAO, though accurate, incurs the highest runtimes due to its sequential sparsity tuning (e.g., 8035.57 seconds for the HN case). In contrast, QC-BAO requires only 149.75 seconds, offering better solution quality in roughly 2% of the time.

Compared to GS-BAO, a fast heuristic with reduced accuracy, QC-BAO demonstrates substantial improvement with only modest overhead. In the lung case, QC-BAO lowers the objective function value from 12.53 to 6.62, with a runtime increase of just 23 seconds. These results highlight QC-BAO's strength in efficiently navigating the non-convex search space while maintaining scalability and performance.

The clinical plans, used as baselines, involve only spot intensity optimization with fixed beam angles, and hence, their runtimes are the shortest. However, the lack of beam angle optimization in clinical plans limits their dosimetric quality and the quality of the objective function value, which QC-BAO improves.

### *3.3 Stochasticity and robustness of QC-BAO*

All BAO methods, except the clinical baseline, require tuning of one additional hyperparameter that governs the strength of the beam angle selection mechanism. For QC-BAO and classical MIP, this is the weight $\mu_2$ in Eq. (4), which penalizes deviation from the desired number of active beams. In our experiments, $\mu_2$ is tuned over the set {0.01, 0.1, 1, 10, 50}.

Both QC-BAO and classical MIP involve stochastic solvers leading to variability across runs. To account for this, we perform five independent runs of each method per test case. Table 1 reports results for the best-performing run (lowest objective function value), while Table 2 summarizes the mean and standard deviation of runtime and objective value over all runs. Furthermore, the average runtime of QC-BAO per run is significantly lower than that of AG-BAO, even when accounting for all five runs. For instance, the total runtime of five QC-BAO runs for the abdomen case is approximately 3280 seconds, still markedly lower than AG-BAO's 9610 seconds. Thus, despite requiring multiple runs, QC-BAO offers faster convergence to better solutions with manageable computational overhead.

Table 1: Comparison of objective function values and runtimes for each method across three clinical test cases (HN, abdomen, and lung). The best generated objective function values are highlighted in **bold**.

| Test case | Quantity | Clinical | GS-BAO | AG-BAO | Classical MIP | QC-BAO |
|---|---|---|---|---|---|---|
| HN | Time (secs) | 40.73 | 289.35 | 8035.57 | 185.42 | 149.75 |
| | Obj fn val | 4.42 | 3.94 | 3.44 | 3.59 | **3.29** |
| Abdomen | Time (secs) | 57.00 | 533.77 | 9610.17 | 1068.7 | 653.03 |
| | Obj fn val | 0.33 | 0.16 | 0.14 | 0.16 | **0.13** |
| Lung | Time (secs) | 140.98 | 599.48 | 14657.95 | 516.10 | 622.34 |
| | Obj fn val | 7.50 | 12.53 | 7.22 | 7.15 | **6.62** |

Table 2: Average computation time (± standard deviation) and average objective function value (± standard deviation) over five runs for classical MIP and QC-BAO.

| Test case | Quantity | Classical MIP | QC-BAO |
|---|---|---|---|
| HN | Time (secs) | 183.2±5.13 | 149.2±1.31 |
| | Obj fn val | 3.83±0.17 | 3.39±0.08 |
| Abdomen | Time (secs) | 1075.21±11.67 | 656.89±6.58 |
| | Obj fn val | 0.16±0.008 | 0.14±0.01 |
| Lung | Time (secs) | 512.44±9.19 | 620.74±2.11 |
| | Obj fn val | 7.23±0.18 | 6.81±0.14 |

*3.4 Selection of beam angles*

In all three test cases, HN, abdomen, and lung, the QC-BAO method selects beam angles that are distinct from both the clinically approved angles and those generated by other beam angle selection methods, highlighting its ability to explore alternative and potentially more effective configurations. In the HN case, QC-BAO employs one 0° couch angle and three 30° angles, with gantry angles distributed across 195°, 60°, 135°, and 345°, deviating from the symmetrical angles used clinically. For the abdomen case, QC-BAO introduces greater variation in couch angles (0°, 30°, 60°) compared to the uniform 0° in the clinical setup, and utilizes non-traditional gantry angles such as 315°, 180°, and 330°. In the lung case, QC-BAO generates the configurations of (0º, 0º, 60º) for couch and (90º, 180º, 345º) for gantry, showing a strategic spatial spread of gantry angles to reduce the dose to OAR.

*3.5 Comparison of plan quality*

The proposed QC-BAO method demonstrates consistently improved treatment plan quality across all three clinical test cases. In terms of target coverage and conformality, QC-BAO yields the highest or tied-best CI in each case. Specifically, in the HN and abdomen cases, it matches AG-BAO with a CI of 0.69 and 0.93 respectively, outperforming both GS-BAO and clinical plans. In the lung case, QC-BAO matches the CI of the clinical plan (0.89), while exceeding those of GS-BAO and AG-BAO.

QC-BAO also demonstrates advantages in OAR sparing. In the HN case, QC-BAO reduces the maximum and mean doses to the right parotid to exceptionally low values (0.08 Gy and 0.0007 Gy, respectively), and achieves the lowest mean dose to the oropharynx (4.49 Gy). In the abdomen case, it achieves the lowest maximum dose to the large bowel (12.19 Gy) and ties AG-BAO for the lowest mean dose to the left bowel (0.29 Gy). While AG-BAO performs better in terms of spinal cord sparing in this case, no single method, including AG-BAO, outperforms QC-BAO across all evaluated metrics. In the lung case, QC-BAO results in the lowest doses to OAR, including the lung ($D_{mean}$ 2.78 Gy), heart ($D_{mean}$ 0.57 Gy) and esophagus ($D_{mean}$ 1.35 Gy). Additionally, QC-BAO yields the lowest mean dose to the body in all three test cases, indicating more efficient and localized dose delivery.

These improvements are further illustrated by the DVH curves shown in Figures 1-3, which confirm the improved OAR sparing, and sharper target dose conformity achieved by QC-BAO. Overall, the results highlight QC-BAO's consistent ability to enhance target dose uniformity while reducing OAR dose exposure.

Table 3: Comparison of the output for the HN case. $V_{30}$ denotes the percentage of OAR volume that receives at least 30 Gy dose. The best values in each row are highlighted in **bold**.

| Structure | Quantity | Clinical | GS-BAO | AG-BAO | QC-BAO |
|---|---|---|---|---|---|
| | Selected couch angles | (0º, 0º, 0º, 0º) | (30º, 30º, 30º, 60º) | (0º, 30º, 30º, 60º) | (0º, 30º, 30º, 30º) |
| | Selected gantry angles | (45º, 135º, 225º, 315º) | (60º, 75º, 180º, 270º) | (270º, 180º, 315º, 300º) | (195º, 60º, 135º, 345º) |
| CTV | $D_{max}$ | 119.33% | 118.40% | **116.10%** | 118.42% |
| | CI | 0.62 | 0.64 | 0.69 | **0.69** |
| R Parotid ($V_{30}$<50%) | $D_{max}$ (Gy) | 4.33 | 10.13 | 1.56 | **0.08** |
| | $D_{mean}$ (Gy) | 0.19 | 3.20 | 0.02 | **0.0007** |
| Oropharynx ($D_{max}$<20 Gy) | $D_{max}$ (Gy) | 35.79 | 34.17 | **32.61** | 34.88 |
| | $D_{mean}$ (Gy) | 4.69 | 5.08 | 4.58 | **4.49** |
| Body | $D_{mean}$ (Gy) | 0.19 | 0.21 | 0.19 | **0.14** |

Table 4: Comparison of the output for the abdomen case. $V_{25}$ denotes the percentage of OAR volume that receives at least 25 Gy dose. The best values in each row are highlighted in **bold**.

| Structure | Quantity | Clinical | GS-BAO | AG-BAO | QC-BAO |
|---|---|---|---|---|---|
| | Selected couch angles | (0º, 0º, 0º) | (0º, 60º, 60º) | (30º, 60º, 60º) | (0º, 30º, 60º) |
| | Selected gantry angles | (0º, 120º, 240º) | (255º, 300º, 315º) | (315º, 150º, 300º) | (315º, 180º, 330º) |
| CTV | $D_{max}$ | 115.93% | 113.32% | 114.74% | **111.35%** |
| | CI | 0.89 | 0.92 | 0.93 | **0.93** |
| L bowel ($D_{max}$<38 Gy, $V_{25}$<12%) | $D_{max}$ (Gy) | 20.38 | 14.86 | 13.86 | **12.19** |
| | $D_{mean}$ (Gy) | 1.02 | 0.39 | 0.29 | **0.29** |
| Spinal cord ($D_{max}$<25 Gy) | $D_{max}$ (Gy) | 15.26 | 6.05 | **0.15** | 1.26 |
| | $D_{mean}$ (Gy) | 3.84 | 1.09 | **0.009** | 0.03 |
| Body | $D_{mean}$ (Gy) | 1.11 | 1.28 | 1.16 | **1.03** |

Table 5: Comparison of the output for the lung case. $V_{27}$ denotes the percentage of OAR volume that receives at least 27 Gy dose. The best values in each row are highlighted in **bold**.

| Structure | Quantity | Clinical | GS-BAO | AG-BAO | QC-BAO |
|---|---|---|---|---|---|
| | Selected couch angles | (0º, 0º, 0º) | (0º, 0º, 30º) | (0º, 30º, 30º) | (0º, 0º, 60º) |
| | Selected gantry angles | (0º, 120º, 240º) | (255º, 270º, 280º) | (0º, 255º, 300º) | (90º, 180º, 345º) |
| CTV | $D_{max}$ | **120.80%** | 126.03% | 126.11% | 121.13% |
| | CI | 0.89 | 0.76 | 0.86 | **0.89** |
| Lung ($D_{mean}$<18 Gy, $V_{12}$<30%) | $D_{mean}$ (Gy) | 3.36 | 4.80 | 3.03 | **2.78** |
| Heart ($V_{27}$<60%) | $D_{mean}$ (Gy) | 0.79 | 1.47 | 0.72 | **0.57** |
| | $V_{27}$ | 1.00% | 1.71% | 1.07% | **0.78%** |
| Esophagus ($D_{mean}$<20 Gy) | $D_{mean}$ (Gy) | 3.25 | 15.75 | 6.39 | **1.35** |
| Body | $D_{mean}$ (Gy) | 2.14 | 2.98 | 2.15 | **1.91** |

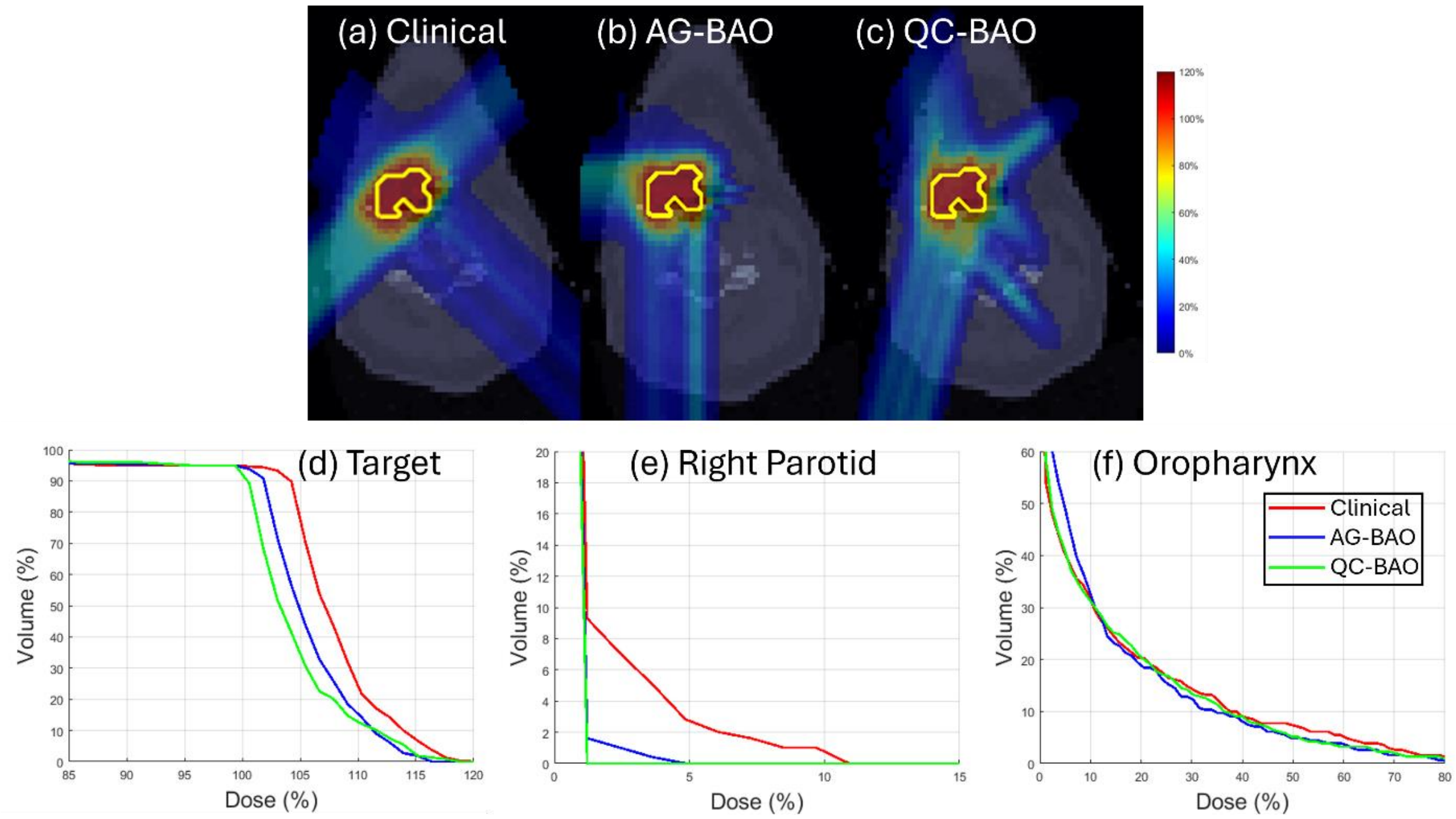

Figure 1 (HN): (a)-(c) Dose plots, (d) DVH plot for target, (e)-(f) DVH plots for OAR

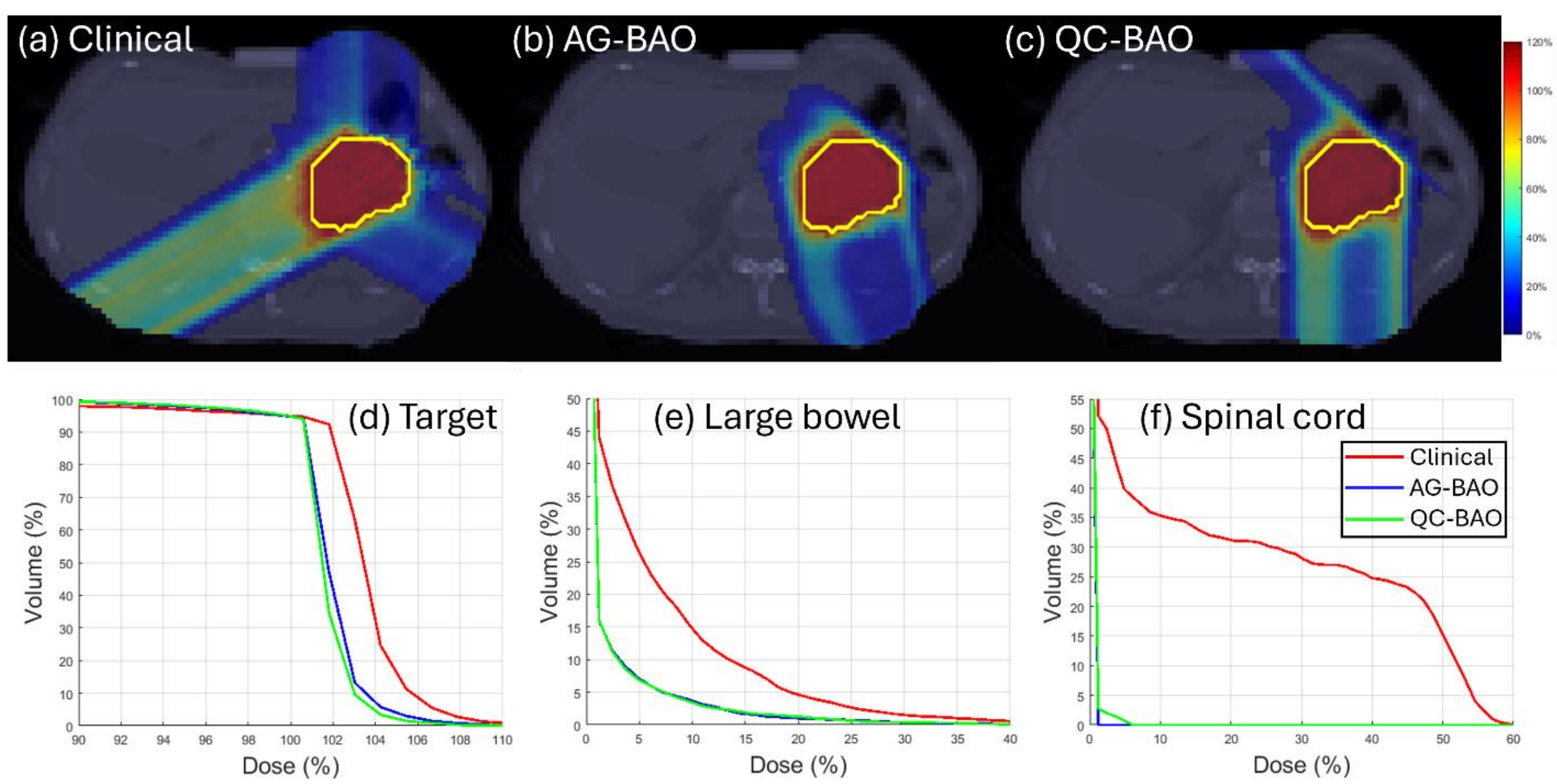

Figure 2 (Abdomen): (a)-(c) Dose plots, (d) DVH plot for target, (e)-(f) DVH plots for OAR

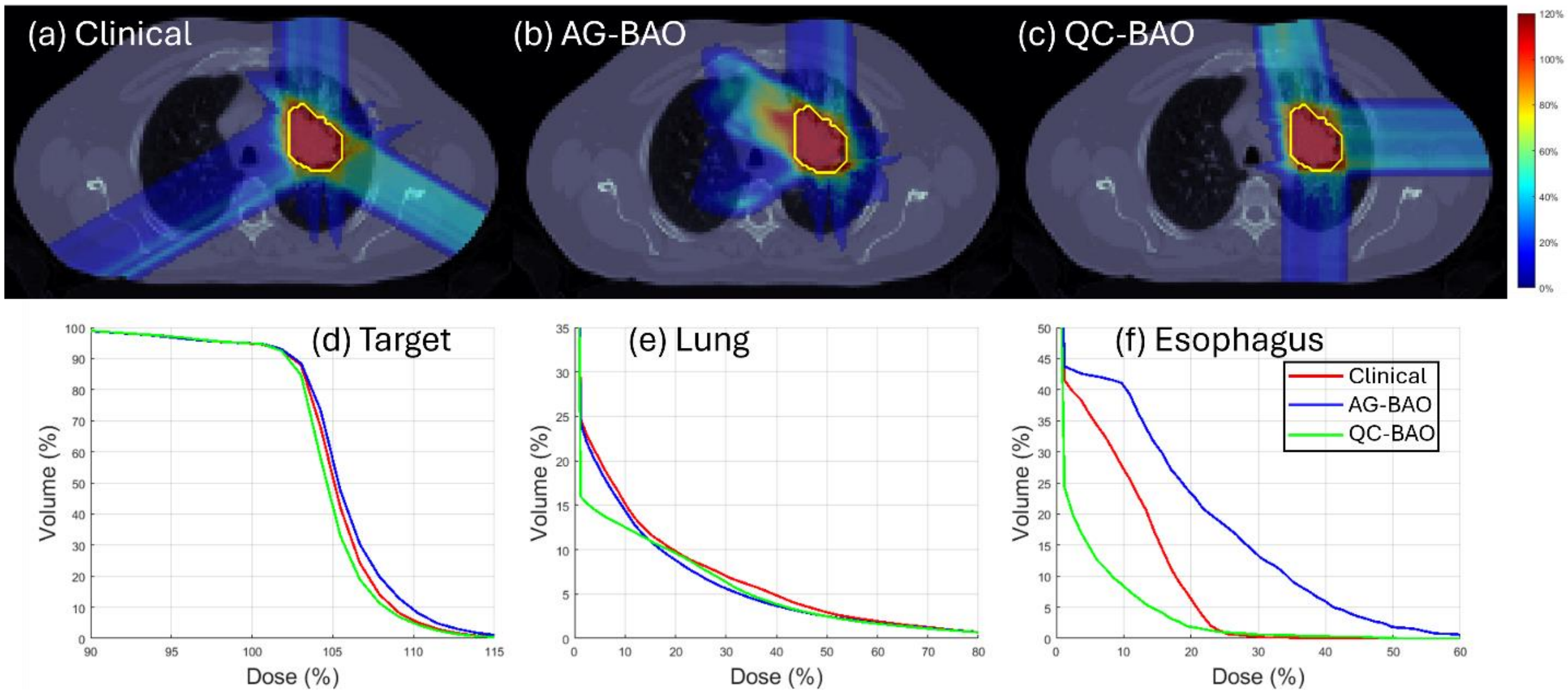

Figure 3 (Lung): (a)-(c) Dose plots, (d) DVH plot for target, (e)-(f) DVH plots for OAR

**5. Discussion**

This work introduces QC-BAO, a novel hybrid classical-quantum approach for solving the MIP formulation of BAO. The results demonstrate that QC-BAO achieves lower objective function values and improved treatment plan quality compared to clinical baselines, classical MIP solvers, and state-of-the-art heuristics such as GS-BAO and AG-BAO, all while maintaining favorable computational efficiency. These improvements highlight the potential of QC-BAO for future clinical deployment.

Importantly, while QC-BAO is designed to be compatible with quantum hardware, the current implementation is executed on a quantum simulator using classical hardware. Although this setup does not leverage quantum parallelism, it retains benefits such as improved exploration of the non-convex binary search space. This allows for the assessment of quantum-inspired strategies in the absence of readily available quantum devices.

While the current approach significantly enhances computational efficiency, further refinements can improve performance. Developing problem-specific quantum algorithms tailored for BAO could optimize solution accuracy and further accelerate computation. Additionally, replacing the hybrid classical-quantum approach for solving the QUBO subproblem with a direct MIP solver [49] could further reduce computational overhead, increasing efficiency and scalability for real-world applications.

QC-BAO also complements a recent study [50] in quantum RT optimization, which applies QUBO to optimize spot intensities. While our work focuses on the upstream task of beam angle selection, these approaches together point toward the potential development of a fully quantum-enhanced optimization pipeline for RT planning.

Beyond BAO, quantum computing holds promise for addressing other complex combinatorial optimization problems in RT that involve discrete decisions. In proton therapy, QC can improve energy layer optimization [51], leading to better dose distributions. Similarly, in spatially fractionated radiation therapy, QC can optimize LATTICE therapy peak placement [52] to enhance tumor control. These problems inherently involve binary decision-making, making them well-suited for QC-based MIP formulations.

Other recent works have demonstrated the growing role of QC across various RT applications. Quantum annealing has been explored for dose distribution optimization [53], significantly reducing computational time while maintaining treatment effectiveness. Additionally, quantum-inspired machine learning models [54-56] have improved tumor segmentation and prediction, increasing accuracy and efficiency. These developments show the expanding role of QC in transforming radiation therapy optimization.

Further bolstering QC's potential are advancements in quantum hardware, including improved qubit stability, error correction techniques, and increased qubit counts. Innovations in superconducting qubits (IBM [28], Google [57]), trapped ions (IonQ [58]), and quantum annealers (D-Wave [27]) have enhanced computational capabilities, making QC more practical for real-world applications. With more powerful quantum processors, RT optimization tasks that were previously infeasible due to computational complexity can now be tackled more efficiently, potentially enabling more precise and personalized treatment plans.

To integrate QC-inspired BAO into clinical practice, rigorous validation is essential. Experimental validation [59,60] through phantom studies can provide quantitative performance metrics, paving the way for controlled clinical trials to ensure safety, efficacy, and practical implementation in patient treatments. From a practical standpoint, the use of non-coplanar beams, as implemented in the current version of QC-BAO, introduces challenges such as increased setup complexity, extended treatment times, and potential collision risks. The present mathematical framework assumes geometric feasibility and does not incorporate these clinical constraints. Future work will focus on embedding machine-specific and deliverability-aware constraints into the optimization process to enhance the clinical viability of the proposed beam configurations.

Moreover, BAO in non-coplanar field setups may offer only marginal improvements if beam directions are already well-chosen. Nevertheless, QC-BAO provides a flexible and extensible framework. A particularly promising direction for future research is to adapt the method for arc-based or VMAT planning, where angle selection has a more pronounced effect on dosimetric outcomes and cannot easily be compensated by spot intensity optimization.

By integrating quantum computing into the BAO framework, this work addresses key computational challenges in RT treatment planning. The proposed approach improves the efficiency of beam angle selection, contributing to enhanced RT outcomes. Future research will focus on refining quantum algorithms, exploring direct MIP solutions using QC, and expanding QC applications to solve other critical problems in RT.

## 6. Conclusions

This work proposes a quantum optimization-based approach to solve mixed integer programming formulation of BAO to identify a near-optimal combination of beam angles from a set of available non-coplanar angles. The proposed model enhances both the objective function value of the MIP problem and the overall quality of the dose plan. Additionally, the integration of a quantum optimization techniques aids in efficiently solving the binary quadratic optimization problem.

## Acknowledgments

This research is partially supported by the NIH grants No. R37CA250921, R01CA261964, and a KUCC physicist-scientist recruiting grant.

## Conflict of Interest Statement

None.